\documentstyle[12pt]{article}
\begin{document}
\title{The gauge boson contributions to the radiatively corrected mass
of the scalar Higgs boson in the minimal supersymmetric standard model}
\author{Seung Woo Ham and Sun Kun Oh
        \\{\it Department of Physics, Kon-Kuk University,} \\
          {\it Seoul 143-701, Korea}
\\
\\
        Bjong Ro Kim
\\
{\it III. Physikalisches. Institut. A, RWTH Aachen,} \\
{\it D52056 Aachen, Germany}
\\
\\
}
\date{}
\maketitle
\thispagestyle{empty}
\begin{abstract}
We derive analytic formulas for the radiatively corrected mass of the scalar
Higgs boson in the framework of the minimal supersymmetric standard model (MSSM).
Since the scalar-top-quark mass in our analysis include terms proportional
to the gauge couplings in the 1-loop effective potential, the radiatively
corrected mass of the scalar Higgs boson partially contains the gauge boson
contributions.
At the 1-loop level, the upper bound on the lighter scalar Higgs boson mass
can be increased about 20 GeV in favor of the partial contributions of the
gauge bosons.
Thus the improved absolute upper bound on the lighter scalar Higgs boson mass
is about 150 GeV.
\end{abstract}
\vfil

\section{Introduction}

\hspace*{6.mm}
The electroweak gauge boson masses $(m_W, m_Z)$ in the standard model
(SM) are generated through the Higgs mechanism.
Then the SM predicts a new scalar particle, the Higgs particle.
But an evidence for the existence of the Higgs particle has not been
experimentally observed up to now.
Therefore a study for the Higgs particle is one of the most important
issues in elementary particle physics at the present.
In the SM only one Higgs doublet is required to give masses to quarks
and leptons.
Therefore there is just one scalar Higgs boson in the SM.
The lower bound of about 60 GeV on the Higgs boson mass of the SM is
estimated from the negative experiment results of LEP 1.
The upper bound on the Higgs boson mass of the SM is known as almost
free parameter.
On the other hand, supersymmetric models impose a strong constraint on the
upper bound on the Higgs boson mass.
In supersymmetric models, at least two Higgs doublets are required to give
masses to fermions via spontaneous symmetry breaking.
Hence, several Higgs bosons are present in supersymmetric models.

One of the most widely studied supersymmetric models is the minimal
supersymmetric standard model (MSSM).
The MSSM is the simplest supersymmetric extension of the SM.
The MSSM offers a solution to the gauge hiearchy problem in a technical way.
Its Higgs sector consists of two Higgs doublets, $H_1$ and $H_2$.
The hypercharges of two Higgs doublets are Y = $- 1$ for $H_1$ and Y = 1
for $H_2$, respectively.
In the MSSM, down-type quarks and leptons are generated in terms of the
vacuum expectation value $(v_1)$ of $H_1$ while up-type quarks are generated
in terms of the vacuum expectation value $(v_2)$ of $H_2$.
The Higgs spectra of the MSSM consists of two neutral scalar Higgs boson
$(h, H)$, one neutral pseudoscalar Higgs boson $(A)$, a pair of charged Higgs
bosons $(H^{\pm})$.

At the MSSM the tree level mass of the Higgs boson depends on only two free
parameters.
Usually, one of two parameters is chosen to be the ratio of the vacuum
expectation values of two Higgs doublets, $\tan \beta$, and the other be
either the lighter scalar Higgs boson mass or the pseudoscalar Higgs boson
mass.
According to the tree level potential, the mass of the lighter scalar Higgs
boson has to be smaller than both the Z boson mass and the pseudoscalar Higgs
boson mass.
Also the MSSM predicts at the tree level that the mass of the charged Higgs
boson must be larger than the $W$ boson mass.
At the tree level, another distinct feature is that the Higgs boson mass
posesses a symmetric property under interchange
$\cos \beta \leftrightarrow \sin \beta$.

It has been recently pointed out that radiative corrections give a positive
contribution to the tree level mass of the scalar Higgs boson.
Especially the dominant contribution to the tree level Higgs boson mass comes
from radiative correctios due to the top-quark and scalar-top-quark loops.
The radiatively corrected mass of the lighter scalar Higgs boson can be larger
than both the Z boson mass and the radiatively corrected mass of the
pseudoscalar Higgs boson [1].
Furthermore, the radiatively corrected mass of the charged Higgs boson can be
smaller than $W$ boson mass [2].
This means that radiative corrections to the tree level mass of the charged
Higgs boson can be negative contributions.
When radiative corrections to the tree level mass of the Higgs boson are
included, the symmetric property of the neutral Higgs boson mass under
interchange $\cos \beta \leftrightarrow \sin \beta$ is broken [3].

In this paper, we calculate the radiatively corrected mass of the neutral
Higgs boson in the framework of the MSSM.
Especially, we consider the gauge boson contributions to the neutral Higgs
boson mass at the 1-loop level.
That is, we take the scalar-top-quark mass which contains terms proportional
to the gauge couplings in the 1-loop effective potential.
Then we derive analytic formulas for the radiatively corrected mass of the
neutral Higgs boson.
The radiatively corrected mass of the pseudoscalar Higgs boson does not changed
by the inclusion of terms proportional to the gauge couplings in the
scalar-top-quark mass.
On the other hand, the radiatively corrected mass of the scalar Higgs boson can
be improved by the inclusion of terms proportional to the gauge couplings in the
scalar-top-quark mass.
At the 1-loop level, the improved mass of the scalar Higgs boson partially
contains the gauge boson contributions.

\section{The Higgs boson mass spectra}

\hspace*{6.mm}
In the MSSM, the tree level Higgs potential including the soft supersymmetry
(SUSY) breaking terms is given by
\begin{eqnarray}
 V_{\rm tree} &=& m_1^2 |H_1|^2 + m_2^2 |H_2|^2
                - m_{3}^2 ( H_1 H_2 + {\rm H.c.} ) \cr
              & &\mbox{} + {1\over 8}(g_1^2 + g_2^2)( |H_1|^2-|H_2|^2 )^2
                + {g_2\over 2}|H_1^* H_2|^2 \ ,
\end{eqnarray}
where $g_1$ and $g_2$ are the U(1) and SU(2) gauge coupling constants,
respectively, and $H_1 H_2 = H_1^0 H_2^0-H^-H^+$.
We assume that $m_i (i = 1, 2, 3) > 0$.

The 1-loop effective potential including the field-dependent mass-squared
matrix ${\cal M}^2$ of the scalar-top-quark [4] is
\begin{equation}
        V_{\rm{1-loop}} = \frac{1}{64\pi^2} Str {\cal M}^4 \left(
           \log \frac{{\cal M}^2}{\Lambda^2} - \frac{3}{2} \right) \ ,
\end{equation}
where the supertrace is defined as
\begin{equation}
        Str f({\cal M}^2) = \sum_i (-1)^{2J_i} (2J_i+1) f(M_i^2) \ .
\end{equation}
Their spins are $J_i = 1/2$ for the top-quark and $J_i = 0$ for the
scalar-top-quark, respectively.
The arbitrary scale $\Lambda$ is taken to be $M_{\rm{SUSY}}$ = 1 TeV.
Thus the 1-loop effective potential due to the top-quark and scalar-top-quark
loops is expressed by
\begin{eqnarray}
    V_{\rm{1-loop}} & = &\frac{3}{32\pi^2} M_{\tilde{t_i}}^4
   \left (\log {M_{\tilde{t_i}}^2 \over \Lambda^2} - {3\over 2} \right )
        - \frac{3}{16\pi^2} M_t^4
        \left (\log {M_t^2 \over \Lambda^2}  - {3\over 2} \right )  \ ,
\end{eqnarray}
where $M_{\tilde{t_i}} (i = 1, 2)$ and $M_t$ are the field-dependent
scalar-top-quark masses and the field-dependent top-quark mass, respectively.
The field-dependent mass-squared of the top-quark is given by
\begin{equation}
        M_t^2 = h_t^2 |H_2^0|^2     \ .
\end{equation}
The field-dependent mass-squared matrix of the scalar-top-quark is given as a
$2 \times 2$ Hermitian matrix.
The eigenvalues of the Hermitian matrix are the field-dependent mass-squareds
of the scalar-top-quark,
\begin{eqnarray}
     M_{\tilde{t_1}, \tilde{t_2}}^2 & = &
     h_t^2 |H_2^0|^2 + {1 \over 2}(m_Q^2 + m_T^2)
     + {1 \over 8}(g_1^2 + g_2^2)(|H_1^0|^2 - |H_2^0|^2)
     \mp \sqrt{R(H_1^0, H_2^0)} \ ,
\end{eqnarray}
with
\begin{eqnarray}
  R(H_1^0, H_2^0) & = &
  \left [ {1 \over 2}(m_Q^2 - m_T^2)
     +({1 \over 4} g_2^2 - {5 \over 12} g_1^2)
     (|H_1^0|^2 - |H_2^0|^2) \right ]^2 \cr
    & &\mbox{} + h_t^2 | \mu H_1^0 + A_t H_2^{0*} |^2 \ .
\end{eqnarray}
In the above equations, $h_t$ is the top-quark Yukawa coupling,
$A_t$ is the trilinear soft SUSY breaking parameter, and
$m_Q$ and $m_T$ are the soft SUSY breaking scalar-quark masses.
Thus the mass-squared of the top-quark is given as
\begin{equation}
        m_t^2 = (h_t v \sin \beta)^2
\end{equation}
and also the mass-squareds of the scalar-top-quark are given as
\begin{eqnarray}
        m_{\tilde{t}_{1,2}}^2 & = & m_t^2 + {1\over 2}(m_Q^2 + m_T^2)
        + {1\over 4} m_Z^2 \cos 2 \beta            \cr
        &   & \mbox{} \mp
        \left [\left ({1\over 2}(m_Q^2 -m_T^2)
        +({2\over 3} m_W^2 - {5\over 12} m_Z^2) \cos 2 \beta \right )^2
        + m_t^2(A_t + \mu \cot \beta)^2 \right]^{{1\over 2}} \ ,
\end{eqnarray}
where $m_{\tilde{t_1}} \le m_{\tilde{t_2}}$ and $m_Z = (g_1^2 + g_2^2) v^2/2,
m_W = (g_2 v)^2$ for $\sqrt{v_1^2 + v_2^2}$ = 175 GeV.

The full scalar Higgs potential up to the 1-loop level may be decomposed as
\begin{equation}
       V = V_{\rm{tree}} + V_{\rm{1-loop}} \ .
\end{equation}
Next let us calculate the mass-squared matrix of the scalar Higgs boson.
The elements of the mass-squared matrix for the neutral Higgs boson are given
by
\begin{equation}
        M_{ij} = \left . \left(
        {\partial^2 V \over \partial \phi_i \partial \phi_j}
                   \right)
        \right| _{<H^0_1> = v_1,\, <H^0_2> = v_2} \ ,
\end{equation}
where $\phi_i$ are the conventional notations for the real and
imaginary parts of the Higgs fields.
The soft SUSY breaking parameters $m_1$ and $m_2$ can be eliminated by the
minimization conditions,
\begin{equation}
{\partial <V> \over \partial v_i} = 0 \ .
\end{equation}
The real parts of two neutral Higgs fields in $M^1_{ij}$ lead to the
mass-squared matrix of the scalar Higgs boson.
The exact expression for the elements of the mass-squared matrix $M^S$
is obtained as
\begin{eqnarray}
  M_{S_{11}} & = & (m_A \sin \beta)^2 + (m_Z \cos \beta)^2 + \Delta_{11}  \ , \cr
  M_{S_{22}} & = & (m_A \cos \beta)^2 + (m_Z \sin \beta)^2 + \Delta_{22}  \ , \cr
  M_{S_{12}} & = & - \mbox{}(m_A^2 + m_Z^2) \sin \beta \cos \beta + \Delta_{12} \ ,
\end{eqnarray}
with
\begin{eqnarray}
\Delta_{11} & = & {3 \over 8 \pi^2} \Delta_1^2
        {\rm g} (m_{\tilde{t_1}}^2,m_{\tilde{t_2}}^2)
        +{3 m_Z^4 \cos^2 \beta \over 128 \pi^2 v^2}
        \log {m_{\tilde{t_1}}^2 m_{\tilde{t_2}}^2\over \Lambda^4}  \cr
        & &\mbox{} - {3 \over 16 \pi^2 v^2}
        \left ({4 \over 3} m_W^2 - {5 \over 6} m_Z^2 \right )^2 \cos^2 \beta
        {\rm f}(m_{\tilde{t_1}}^2,m_{\tilde{t_2}}^2)      \cr
        & &\mbox{} + {3 \over 16 \pi^2 v} m_Z^2 \cos \beta
  \left ({\Delta_1 \over m_{\tilde{t_1}}^2 - m_{\tilde{t_2}}^2} \right )
        \log {m_{\tilde{t_1}}^2 \over m_{\tilde{t_2}}^2}   \ ,\cr
        & &  \cr
\Delta_{22} & = & {3 \over 8 \pi^2} \Delta_2^2
        {\rm g} (m_{\tilde{t_1}}^2,m_{\tilde{t_2}}^2)
        - {3 m_t^4 \over 4 \pi^2 v^2 \sin^2 \beta}
        \log {m_t^2 \over \Lambda^2}  \cr
        & &\mbox{} - {3 \over 16 \pi^2 v^2}
        ({4 \over 3} m_W^2 - {5 \over 6} m_Z^2)^2 \sin^2 \beta
        {\rm f} (m_{\tilde{t_1}}^2,m_{\tilde{t_2}}^2)       \cr
        & &\mbox{} + {3 \over 16 \pi^2 v}
        \left ({4 m_t^2 \over \sin \beta} - m_Z^2 \sin \beta \right )
   \left ({\Delta_2 \over m_{\tilde{t_1}}^2 - m_{\tilde{t_2}}^2} \right )
        \log {m_{\tilde{t_1}}^2 \over m_{\tilde{t_2}}^2}  \cr
        & &\mbox{} + {3 \over 32 \pi^2 v^2}
        \left ({2 m_t^2 \over \sin \beta}
        - {1 \over 2} m_Z^2 \sin \beta \right )^2
        \log {m_{\tilde{t_1}}^2 m_{\tilde{t_2}}^2 \over \Lambda^4}  \ , \cr
        & &  \cr
\Delta_{12} & = & {3 \over 8 \pi^2} \Delta_1 \Delta_2
        {\rm g} (m_{\tilde{t_1}}^2,m_{\tilde{t_2}}^2)
        + {3 m_Z^2 \sin 2 \beta \over 256 \pi^2 v^2}
        \left( {4 m_t^2 \over \sin^2 \beta } -m_Z^2 \right)
        \log {m_{\tilde{t_1}}^2 m_{\tilde{t_2}}^2 \over \Lambda^4}  \cr
        & &\mbox{} + {3 \over 32 \pi^2 v} \left [m_Z^2 \cos \beta \Delta_2
        + \left ({4 m_t^2 \over \sin^2 \beta} - m_Z^2 \right )
        \sin \beta \Delta_1 \right]
        {1 \over (m_{\tilde{t_1}}^2 - m_{\tilde{t_2}}^2)}
        \log {m_{\tilde{t_1}}^2 \over m_{\tilde{t_2}}^2}  \cr
        & &\mbox{} + {3 \over 32 \pi^2 v^2}
        \left ({4 \over 3} m_W^2 - {5 \over 6} m_Z^2 \right )^2 \sin 2 \beta
        {\rm f} (m_{\tilde{t_1}}^2,m_{\tilde{t_2}}^2)       \ ,
\end{eqnarray}
as well as
\begin{eqnarray}
        \Delta_1  & = & {m_t^2 \mu \over v \sin \beta}
        (A_t + \mu \cot \beta)             \cr
        & &\mbox{} + {\cos \beta \over 2 v}
        \left [ (m_Q^2 - m_T^2) + ({4\over 3} m_W^2
        - {5\over 6} m_Z^2) \cos 2 \beta \right]
        ({4\over 3} m_W^2 - {5\over 6} m_Z^2)  \ , \cr
        \Delta_2  &=& {m_t^2 A_t\over v \sin \beta}
        (A_t + \mu \cot \beta)             \cr
        & &\mbox{} - {\sin \beta \over 2 v}
        \left [ (m_Q^2 - m_T^2) + ({4\over 3} m_W^2
        - {5\over 6} m_Z^2) \cos 2 \beta \right]
        ({4\over 3} m_W^2 - {5\over 6} m_Z^2)  \ ,
\end{eqnarray}
and also
\begin{eqnarray}
      {\rm f} (m_{\tilde{t_1}}^2,m_{\tilde{t_2}}^2) & = &
      {1 \over (m_{\tilde{t_2}}^2-m_{\tilde{t_1}}^2)}
      \left[  m_{\tilde{t_1}}^2 \log {m_{\tilde{t_1}}^2
      \over \Lambda^2} -m_{\tilde{t_2}}^2
      \log {m_{\tilde{t_2}}^2 \over \Lambda^2}
      \right] + 1  \ , \cr
      {\rm g} (m_{\tilde{t_1}}^2,m_{\tilde{t_2}}^2)
        &=& {1 \over (m_{\tilde{t_1}}^2 - m_{\tilde{t_2}}^2)^3}
        \left [(m_{\tilde{t_1}}^2+m_{\tilde{t_2}}^2)
        \log {m_{\tilde{t_2}}^2 \over m_{\tilde{t_1}}^2}
        - 2 (m_{\tilde{t_2}}^2 - m_{\tilde{t_1}}^2) \right ] \ .
\end{eqnarray}
In the above equation, the radiatively corrected mass-squared of the
pseudoscalar Higgs boson is given by
\begin{equation}
m_A^2 = {m_3^2 \over \sin \beta \cos \beta}
        + {3 m_t^2 A_t \mu \over 16 \pi^2 v^2 \sin^3 \beta \cos \beta}
        {\rm f}(m_{\tilde{t_1}}^2,m_{\tilde{t_2}}^2)       \ ,
\end{equation}
where the first term stands for the mass-squared of the pseudoscalar
Higgs boson at the tree-level.
Even if terms proportional to the gauge couplings in the scalar-top-quark
mass are included, the radiatively corrected mass of the pseudoscalar Higgs
boson does not changed.
The mass-squareds of the scalar Higgs boson are obtained as
\begin{eqnarray}
   m_{h, H}^2 & = & {1 \over 2}
      \left [ Tr M^S \mp \sqrt{(Tr M^S)^2 - 4 det (M^S)} \right ] \ ,
\end{eqnarray}
$m_h \le m_H$.

Now let us set a reasonable region in parameter space in order to obtain
the numerical result for the neutral Higgs boson mass.
The CDF and D0 collaborations, respectively, predict that the top-quark
mass is $176 \pm 8 \pm 10$ and $199^{+19}_{-21} \pm 22$ GeV [5].
Therefore we take the top quark mass of 175 GeV.
The Higgs mixing parameter $\mu$ is the mass parameter introduced in the
superpotential term $\mu H_1 H_2$.
The values of the function ${\rm f} (m_{\tilde{t_1}}^2,m_{\tilde{t_2}}^2)$
are always positive in our parameter setting.
If we set $\mu \cdot A_t$ as a positive value, radiative corrections will be
positively contributed to the tree level mass of the pseudoscalar Higgs boson.
In this paper, we fix $\mu$ as the electroweak scale, 100 GeV.
In our numerical analysis, the ranges of parameters are presented in Table 1.
The upper limit of $\tan \beta$ in Table. 1 is given by equation,
$\tan \beta =m_t/m_b \approx 40$ for $m_t$ = 175 GeV
The upper limit on the tree level mass of the pseudoscalar Higgs boson is
fixed as 950 GeV in our numerical analysis.
Thus $m_3^2$ can be increased up to $150^2$ for $\tan \beta$ = 40.
Experimentally supersymmetric particles are constrained to be heavier than
their SM partners [6].
Therefore we assume that the lower limit on the lighter scalar-top-quark mass
is larger than the top-quark mass.
\begin{center}
Table I. The ranges for the MSSM parameters

\begin{tabular}{ccccc}  \hline \hline
$10^2$  & $\le$  & $m_3^2$       & $\le$  & $150^2$         \cr
2       & $\le$  & $\tan \beta$  & $\le$  & 40              \cr
100     & $\le$  & $m_Q$         & $\le$  & 1000 GeV        \cr
100     & $\le$  & $m_T$         & $\le$  & 1000 GeV        \cr
100     & $\le$  & $A_t$         & $\le$  & 3000 GeV        \cr
\hline \hline
\end{tabular}
\end{center}

Next let us evaluate the numerical result for the upper bound on the neutral
Higgs boson mass including radiative corrections in the MSSM.
In Fig. 1, we plot $m_h$ including the partial contributions of the gauge
bosons at the 1-loop level.
$m_h$ is maximized for the parameter space of $10^2 \le m_3^2 \le 150^2$ GeV,
and $100 \le m_Q (m_T) \le 1000$ GeV.
We find from Fig. 1 that $m_h$ increases as $\tan \beta$ increases.
That is, $m_h$ of the MSSM is always maximized as $\sin \beta \rightarrow 1$.
Fig. 2 shows the upper bound on the radiatively corrected mass of the neutral
Higgs boson, as a function of $\tan \beta$, for $10^2 \le m_3^2 \le 150^2$
GeV, $100 \le m_Q (m_T) \le 1000$ GeV, and $100 \le A_t \le 3000$ GeV.
Of course the upper bound on the lighter scalar Higgs boson mass at the tree
level is $Z$ boson mass.
The dashed curve exhibits the upper bound on the neutral Higgs boson mass
without the gauge boson contributions.
The solid curve exhibits the upper bound on the neutral Higgs boson mass
with the partial contributions of the gauge bosons.
We find from Fig 2 that $m_A$ approaches $m_H$ as $\tan \beta$ increases.
The absolute upper bounds on $m_h$ without the gauge boson contributions and
with the partial contributions of the gauge bosons is about 130 and 150 GeV
for $\tan \beta$ = 40 GeV, respectively.

\section{Conclusions}

\hspace*{6.mm}
In the MSSM, analytic formulas for the radiatively corrected mass of the
neutral Higgs boson are derived from the 1-loop effective pontial.
In our analysis, the scalar-top-quark mass in the 1-loop effective potential
includes terms proportional to the gauge couplings.
The radiatively corrected mass of the pseudoscalar Higgs boson does not changed
by the inclusion of terms proportional to the gauge couplings in the
scalar-top-quark mass [7].
Radiative corrections to the tree level mass of the pseudoscalar Higgs boson
leads to a positive contribution for $\mu \cdot A_t > 0$ in our parameter space.
Radiative corrections to the pseudoscalar Higgs boson give a positive
contribution of about 90 GeV in its mass.
The formuals for the radiatively corrected mass of the scalar Higgs boson
are different from those of other paper [7].
At the 1-loop level, terms proportional to gauge couplings in the
scalar-top-quark mass give a significant contribution to the scalar Higgs boson
mass.
Especially, the radiatively corrected mass of the lighter scalar Higgs boson
can be increased by about 20 GeV in favor of the partial contributions of the
gauge bosons.
Thus we have obtained the improved absolute upper bound on the lighter scalar
Higgs boson mass of about 150 GeV.

\vskip 0.3 in
\noindent
{\large {\bf Acknowledgements}}

This work is supported in part by the Basic Science Research Institute Program,
Ministry of Education, BSRI-96-2442.
\vskip 0.2 in


\vfil\eject
\noindent

{\bf Figure Captions}

\vskip 0.3 in
\noindent
Fig. 1 : \ The maximum values of $m_h$ at the 1-loop level, as a function of
$A_t$, for $10^2 \le m_3^2 \le 150^2$ GeV, 100 $\le m_Q (m_T) \le$ 1000 GeV.
\vskip 0.2 in
\noindent
Fig. 2 : \ The upper bound on the neutral Higgs boson mass,
as a function of $\tan \beta$, for $10^2 \le m_3^2 \le 150^2$ GeV,
100 $\le A_t \le$ 3000 GeV.
The dashed (solid) curve denote the upper bound on the neutral Higgs boson
mass without (with) the gauge boson contributions.
\vskip 0.2 in
\vfil\eject
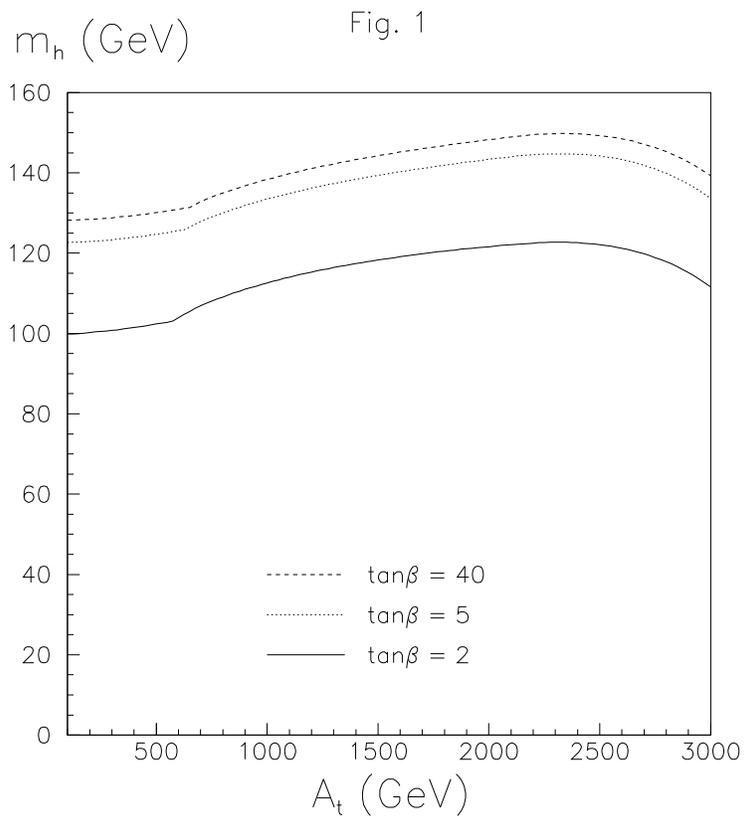
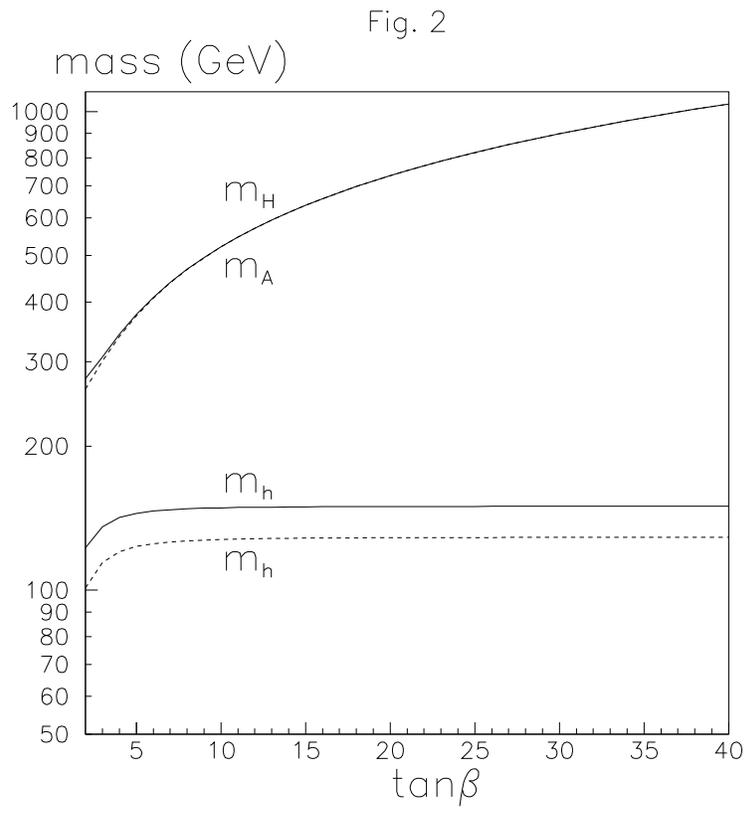
\end{document}